# Doubly and triply coupled nanowire antennas

*Liu Lu[a,b], Lulu Wang[a], Changling Zou[a], Xifeng Ren[a*], Chunhua Dong[a], Fangwen Sun[a], Guoping Guo[a], Guangcan Guo[a]*

[a] Key Laboratory of Quantum Information, University of Science and Technology of China, Hefei 230026, China

[b] School of Mechanical Engineering, Jiangsu University, Zhenjiang 212013, China

*Nanoantenna is one of the most important optical components for light harvesting. In this study, we show experimental evidence of interactions between coupled nanowires by comparing the fluorescence properties of quantum dots on both non-coupled (single) and coupled (double and triple) nanowire arrays. The fluorescence intensities are dependent on the excitation polarization, and most of the emissions are polarized perpendicular to the long axis of the nanowires. It is interesting that both the excitation polarization dependent enhancement and the fluorescence polarization effect are more pronounced for coupled nanowires. Our theoretical analysis indicates the above phenomena can be ascribed to the dipolar plasmon from the nanowire antennas. Our investigations demonstrate a potential method to control the polarization of emitters using coupled nanowire arrays.*

1. Introduction

In the visible and near infrared (NIR) spectra, the electromagnetic field associated with surface plasmons can be confined in a nanoscale volume beyond the diffraction limit. This extreme confinement strongly modifies the photonic local density of the states and tailors the emission properties of emitters. Since the pioneering work of Nie et al. demonstrated that the enhanced Raman scattering of silver nanoparticle clusters was stronger than that of a single nanoparticle,[1] there has been a growing interest in coupled metallic nanostructures.[2-6] Compared to a single nanoparticle, coupled nanoparticles exhibit many unique optical properties, such as providing "hot spots" in the electromagnetic field, and lighting up multipolar surface plasmon resonance, and the Fano resonance.[6] Coupled nanoparticles can also act as an antenna to direct radiation and thus enhance the free space excitation and collection efficiency.[5-7] When combined with the localized surface plasmon field, these antennas can control and manipulate optical fields on the nanoscale, which enhances the efficiency of photo-detection, light emission and sensing.[7-8] Recently, highly directional emissions from quantum dots with a divergence angle of 12.5 or 4° are obtained by embedding the quantum dots into a gold Yagi-Uda antenna and in silver metallic nanoslit arrays, respectively. [9,10]

---

[*] Electronic mail: renxf@ustc.edu.cn

Of the various types of metallic nanostructures, silver nanowires possess certain unique properties that make them particularly promising, such as supporting both localized and propagating surface plasmons, ease of near-field manipulation, low propagation loss, and scattering plasmons to the far field.[11-15] We recently found that single nanowires can improve both the polarization dependent fluorescence enhancement and the emission polarization control.[16] Similarly to coupled nanoparticles, coupled metallic nanowire arrays have recently attracted inreasing attention. It has been theoretically proposed that parallel silver nanowires can form tunable and robust waveguides, while the local surface plasmon in coupled nanowires can lead to a drastic field enhancement, even away from the resonance.[17]

We provide experimental evidence of the interactions between the nanowires in an array by comparing the fluorescence properties of quantum dots on non-coupled (single) and coupled (double and triple) nanowires. It was found that the fluorescence of quantum dots on coupled nanowires was higher in magnitude relative to a single nanowire. In addition, the polarization dependence of the free space light excitation and emission collection demonstrated the antenna effect of coupled silver nanowires. A detailed theoretical analysis confirmed both these interactions and the antenna effect of the surface plasmons in the coupled nanowires.

## 2. Results and Discussion

The samples in our experiments were created by placing the silver nanowires on a glass substrate and covering them with quantum dots, as illustrated in Figure 1a. The CdSe quantum dots act as an optical emitter in this sample system, whereas the silver nanowires support the surface plasmons. The photoluminescence spectrum of the CdSe quantum dots is shown in Figure 1e. The emission wavelength of the quantum dots was 613 nm with a 30 nm FWHM. In addition, the reported silver nanowires were coated by either 8 nm or 35 nm silica using the Stöber method [21,22] to avoid the emission quenching of the quantum dots. [18-20] As shown in Figure 1c and d, the silica coating was relatively smooth and uniform, with a homogeneous thickness over the surface of all of the nanowires, regardless of nanowire diameter.

Figure 1a shows that silver nanowires are arranged in parallel and close together to allow a comparative study of the surface plasmon on both single and coupled nanowires in a single sample. Figure 2a provides the SEM image of three aligned silver nanowires capped with 8 nm of silica. The diameters of these three nanowires are approximately 350 nm, and they are close together with a gap of approximately 35 nm. This gap is composed of 19 nm of air space and 16 nm of silica. The quantum dots on the nanowires cannot be observed in Figure 2a because of the resolution of the SEM image. Marks a, b and c in the figure correspond to singly, doubly and triply aligned silver nanowires, respectively. Figure 2, panels b and c show a charge-coupled device (CCD) picture of the same sample with a 45 degree clockwise rotation relative to the SEM image. We can observe that the proximal quantum dots are immediately excited when the 532 nm laser focuses on site c, as shown in Figure 2c. In addition, it is interesting that each end of the three nanowires also literally lit up. Note that Figure 2c was obtained using a 610±17 nm bandpass filter in front of the CCD to block the laser light; therefore, the observed bright spots were attributed to the quantum dot fluorescence. This phenomenon indicates that the quantum dot fluorescence either radiates directly into free space or excites the propagating surface plasmons in the nanowires, which subsequently scatter photons at their ends.

It is clear that only the surface plasmon from a single nanowire is excited when the incident light focuses on site a in the silver nanowires (Figure 2a). Interestingly, the coupled nanowires showed different results with the surface plasmon for all adjacent nanowires being simultaneously excited, as demonstrated by sites b and c for the double and triple coupled nanowires, respectively. We investigate the fluorescence intensity of the quantum dots versus the incident light polarization under different conditions. The results are shown in Figure 2d, where $\theta$ is the excitation polarization angle relative to the nanowire length axis. The emission spot is selected using a pinhole in the confocal system to block all other fluorescence. In Figure 2d, we can see that the fluorescence intensity varies periodically for all of the emission spots a-c. The surface plasmon in silver nanowires is known to be dependent on the polarization of the incident light,[23] whereas the aggregated quantum dot fluorescence is independent of this polarization. Therefore, the results in Figure 2d indicate the emitted photons exhibit features of both the fluorophore and plasmon resonance. In addition, the quantum dot and surface plasmon coupling proved to always be accompanied by a several fold enhancement in spontaneous emission,[11,24] which is in agreement with the photoluminescence spectra (Figure S2, Supporting Information). Furthermore, both Figure 2d and S2 show that this coupling efficiency is strongest when the incident light polarizaition is perpendicular to the long axis of the nanowires and lowest when parallel. These results seem contradictory to our recent study on single nanowires, which showed the strongest coupling efficiency for parallel polarized light.[16] This discrepancy is caused by the excitation of the silver nanowire from either the bottom or top of the substrate as proven in our theoretical analysis. We also note that the fluorescence intensity for $\theta = 270°$ is lower than for $\theta = 90°$. We attribute this phenomenon to the quenching of the quantum dots, which occurs when the quantum dots are irradiated for long times or when the spacer layer is too thin.[25] The fluorescence intensity can be fitted using a sine function with an exponential decay to account for this quenching. As shown in Figure 2d, the curves can be well fitted by the following equation:

$$f(\theta) = (A + B\sin^2\theta) \cdot \exp(-\beta \cdot \pi \cdot \theta / 180°)$$,

where β is the quenching parameter. Our results gave a value of 0.02 for the β of quantum dots on a single nanowire, which increased to 0.08 for double or triple nanowires. This phenomenon suggests that the coupled nanowires have a greater influence than single nanowires on the quantum dot fluorescence.

Figure 2d shows that the fluorescence intensities for sites a-c are almost identical when θ is 0°. As θ increases from 0° to 90°, the intensity of the coupled sites increases more quickly than the non-coupled sites. All sites reach their maximum intensity when θ is 90°. The fluorescence intensity for site c was strongest and far exceeded that of site a at θ =90°. We attribute this phenomenon to surface plasmon coupling between the doubly or triply coupled nanowires. As for coupled nanoparticles, the coupling efficiency depended on the excitation light polarization. Moreover, the coupling efficiency is strongest when the excitation is polarized perpendicular to the direction of the nanowire and weakest when parallel. Coupling also leads to a dipolar resonance that drastically increases the local field. The doubly and triply aligned nanowires both form a coupled system that possesses an additional dipolar resonance dependent on the excitation polarization relative to a single nanowire. This point is theoretically verified below. To describe the properties of the coupled systems, we define γ as the ratio $I_{max}/I_{min}=I_{90°}/I_{180°}$. The γ values obtained were 1.5, 5.4 and 11.4 for the emissions at spots a, b and c, respectively, which suggests that the coupling effect becomes stronger upon increasing the number of coupled nanowires.

To obtain deeper insight into the dependence of the surface plasmon coupling on the gap thickness, another doubly coupled nanowire sample with a larger gap was investigated. As shown in Figure 1d, these nanowires were capped with 35 nm silica. The SEM image shown in Figure 3a reveals that the nanowires are tightly packed with a silica gap of 70 nm and a diameter of 380 and 300 nm for the upper and lower nanowires, respectively. CCD images with and without laser irradiation are shown in Figure 3b and c, respectively. The phenomenon observed in Figure 3c is similar to that of the triply aligned nanowires. Figure 3d possesses fluorescence intensities dependent on the excitation polarization, which was detected from the sites a-c marked in Figure 3a. Moreover, the fluorescence intensities fit the following sin function closely:

$$f(\theta) = (A + B\sin^2\theta)$$

Obviously, this fitting accounts for the quantum dot quenching's being avoidable at each of the spots because of the 35-nm-thick silica spacer layer. Therefore, in contrast to the results for an 8-nm silica coating, an exponential decay was not required by the fitting function above. In addition, the γ values for the upper and lower single nanowire is 1.6 and 3.2, respectively. The γ value increased to 4.9 as the detected emission spot focused on the junction of both nanowires. This increased γ magnitude is low relative to that of the previous sample with a thinner spacer, which indicates the plasmon coupling efficiency decreases as the gap thickness increases. This demonstration is further discussed and verified by the following theoretical analysis. We also note that the γ value was different for the upper and lower single nanowires, which can be ascribed to the excitation of the nanowire surface plasmon being related to the ratio between the longitudinal and transverse lengths of the nanowire.

We also analyzed the emission polarization of the quantum dots on the nanowires. By rotating another $\lambda/2$ plate (633 nm) in front of the single photon detector (SPD), we can selectively detect the fluorescence intensity at different polarization. Figure 4 presents the polarized fluorescence intensities from emission spots a and c on the triple aligned nanowires (See Figure 2a), where  is the emission polarization angle with respect to the long axis of the silver nanowire. It can be observed that the collected quantum dot fluorescence was primarily polarized perpendicular to the nanowires. In contrast, when using the same setup, the fluorescence intensity of free-space quantum dots distant from any nanowire is independent of the polarization of the excitation light, and the fluorescence polarization is random, which indicates localized plasmon resonance plays a role in both the excitation and emission processes. The curves for a and c in Figure 4 show a fluorescence maximum and minimum at  =90° and  =180°, respectively. Curves a and c also fit well using the sine equation from Figure 2d accompanied by the exp function describing the quenching effect. Moreover, the $I_{max}/I_{min}$ ratio increases from 4.1 to 10.0 as the fluorescence detection changes from the single nanowire (site a) to the triply aligned nanowires (site c). Obviously, the quantum dot emission was also affected by the polarization dependence of the surface plasmon, and this effect is stronger in the coupled nanowires.

To qualitatively analyze the above experimental results, we numerically simulated the electromagnetic field distribution when the nanowires are illuminated with a Gaussian laser beam. The simulation schematic shown in Figure 5a is identical to the case studied in our experiments: silver nanowires (diameter: 350 nm) coated with silica (thickness: 8 nm) are placed on a silica substrate, and the laser beam is perpendicular to the substrate surface, with its electric field being either perpendicular or parallel to the nanowires' long axis. This simulation was performed using the two-dimensional cross-section of the nanowires to reduce the computational resources required.

Figure 5, panels b-d show the electric field intensity ($|E^2|$) around the nanowires when illuminated with a laser with an electric field parallel to the nanowires. We can clearly see that the metal nanowire acts as an obstacle to the beam, and a portion of the light is reflected, while the rest is diffracted by the nanowires. Increasing the number of nanowires increases the surface area of this "mirror", and the nanowires reflect more light. However, when the electric field is perpendicular to the nanowire, as shown in Figure 5, panels e-g, the laser can excite the local surface plasmon in the silver nanowires, which concentrates the electric field onto the nanowire surface.

In our experiments, the quantum dot excitation was well below saturation; therefore, we can estimate that the emission intensity was determined by the average local electric field intensity, $I_{avg}$, at the quantum dots, assuming they are uniformly coated onto the nanowire surface. The calculated $I_{avg}$ of the single, double and triple nanowires were 0.184, 0.203, and 0.197, respectively, for the parallel polarization, and 1.574, 2.733, and 3.477, respectively, for the perpendicular polarization in all three cases. These results mean the quantum dots surrounding the nanowires could be more efficiently excited by a laser with perpendicular polarization than one with parallel polarization, which agrees with our experimental observations.

To determine the physical mechanism of the local field enhancement for multiple nanowires, we calculated the average electric field intensity versus the wavelength, as shown in Figure 6a. These curves show an incremental increase for multiple nanowires over a wide spectral range. There is a peak in the average intensity spectrum near 532 nm, which indicates a local surface plasmon resonance. To verify this resonance, we plotted the field intensity for a point near the local field hot spot in Figure 5, panels e-g. The expected peaks are clearly shown in Figure 6b. Surface plasmon resonance may be responsible for the local field enhancement of the quantum dot excitation by silver nanowires. Single nanowires possess weaker field intensities than the double or triple nanowires at 532 nm, as observed in Figure 6, panels a and b. Therefore, it can be deduced that the modification of the quantum dot fluorescence by multiple coupled nanowires is stronger than for a single nanowire, which agrees with our experimental results.

However, the question is why coupled nanowires result in stronger local electric field intensities. In Figure 7a, we calculate the average intensity against the gap distance between the two nanowires. Although the cross sectional area barely changes, the average intensity varies greatly. We attribute this phenomenon to the antenna effect. Figure 7b shows that two coupled nanowires form a dipole antenna; charges in the nanowires interact strongly.[26] The emissions from this dipole antenna demonstrate directionality, as shown by Figure 7c. Based on classical electrodynamics, the total emission power of the antenna is $P = P_0^2 \omega^4 / 12\pi\varepsilon_0 c^3$, where P$_0$=q×d is the dipole momentum, q is the charge, and d is the gap between them. The interaction decreases at a rate of $d^{-2}$ as the gap increases, which corresponds to the decrease in the emission power of certain charges. At the same laser illumination, the corresponding local charges will decrease with increasing gap distance, with corresponding decreases in the average intensities. These predictions are verified by comparing Figures 2 and 3. Finally, it can be concluded that the antenna effect of the coupled structures is the basis of our results.

## 3. Conclusions

In conclusion, we compared the optical properties of quantum dots on singly, doubly and triply coupled nanowires. The fluorescence intensities for these three situations were shown to depend on the polarization of the incident light with a sinusoidal correlation, and the fluorescence becomes polarized primarily perpendicular to the nanowires. We attribute these phenomena to the plasmon coupling excitation and emission, respectively. Most importantly, we find that the doubly and triply aligned nanowires produce higher enhancements and more strongly polarized fluorescence than single nanowires when the excitation is perpendicularly polarized to the nanowires. However, for parallel excitation, the single, double and triple nanowires yield similar results. According to our systematic theoretical analysis, the reason for the above phenomena is attributable to the antenna effect of the coupled nanowires. Further calculations demonstrate that increasing the number of nanowires to infinity would not only significantly enhance the fluorescence intensity but also perpendicularly polarize the fluorescence relative to the nanowires. Our investigation provides a potential method for perfectly controlling the polarization of an emitter using coupled nanowire arrays. Because this parallel nanowire array can be created experimentally by chemical means,[23] this kind of plasmonic nanoantenna may become widely used for nano-optics in the future.

## 4. Experimental Section

The silver nanowires were prepared by reducing silver nitrate with ethylene glycol in the presence of poly(vinyl pyrrolidone). The silver nanowires were subsequently capped with silica via the Stöber method.[21,22] The preparation details are shown in the Supporting Information. The thickness of the silica sheath can be readily controlled using the reaction time and precursor solution concentration. Finally, a TEM grid was fixed onto the sample with tape to identify the different positions, and three silver nanowires were then closely arranged in parallel using a tapered fiber tip. The experimental setup for studying the nanowires-quantum dot system presented in Figure 1b was based on a modified confocal microscope. A sample placed on a three-dimensional piezoelectric transition (PZT) stage was excited using a continuous-wave laser at 532 nm, and the polarization was controlled using a $\lambda/2$ plate (work wavelength 532 nm) behind a polarization beam splitter (PBS). The laser was focused on the sample using an $100\times$ objective (NA=0.75), and the fluorescence from the QDs was collected using the same objective and detected by a single photon detector (SPD, SPCM-15, PerkinElmer Optoelectronics, Canada). A green-orange dichroic plate was used to reflect the laser light while transmitting the QD fluorescence (the emission peak was centered at 613 nm at room temperature). Spatial filtering was accomplished using a confocal lens with a pinhole (diameter: 30 um) in its center. This setup can collect the light from an area approximately 1.5 μm in diameter on the sample surface. This area can be artificially selected by moving the front lens in the confocal lens system. A narrow band filter (center wavelength 633 nm) was also used to reflect the excitation light before the SPD. The polarization of the fluorescence was analyzed using another $\lambda/2$ plate (work wavelength 633 nm) combined with a PBS. A CCD camera was used for direct imaging.

**Supporting Information**

The optical measurement procedures, nanowire and nanocrystals synthetic details, and photoluminescence spectra for nanocrystals at different emission spots and excitation polarization of 0, 45 and 90° degrees is available from the Wiley Online Library or from the author.

**Acknowledgments**

This work was funded by the National Basic Research Program of China (Grants No. 2011CBA00200 and 2011CB921200), the Innovation funds from Chinese Academy of Sciences (Grants No. 60921091), the National Natural Science Foundation of China (Grants No.10904137), the Fundamental Research Funds for the Central Universities (Grants No. WK2470000005), and the Research Fund for Advanced Talents of Jiangsu University (Grants No. 1281110026)

**Figure captions**

**Figure 1** (a) The pump laser illuminates the sample from the top, and the quantum dot emissions are collected using the same objective lens. CdSe/ZnS quantum dots and silver nanowires are deposited onto a glass substrate. (b) Schematic of the experimental setup. Both (c) and (d) are the TEM images of the silica capped nanowires with silica thicknesses of 8 nm and 35 nm, respectively. (e) The photoluminescence of the CdSe/ZnS quantum dots.

**Figure 2** (a) SEM and (b, c) CCD images of the triply aligned silver nanowires capped with 8 nm silica. (d) The fluorescence intensity as a function of the excitation laser polarization. The fluorescence curves at sites a (dark diamond), b (purple circle), and c (red square) as shown in Figure (a). The lines in (d) are the fitting curves using equation (1).

**Figure 3** (a) The SEM and (b, c) CCD images of doubly aligned silver nanowires capped with 35 nm silica. (d) The fluorescence intensity as a function of excitation polarization. The fluorescence curves at sites a (full square), b (full circle), and c (full diamond) from the three nanowires in Figure (a). The lines in (d) are the best fitting sinusoidal function.

**Figure 4** The emission polarization of the fluorescence detected at sites a and c (Figure 2a) for nanowires capped with 8 nm silica. The full diamonds (curve a) and circles (curve c) correspond to sites a and c, respectively. These points represent the experimental data, while the lines represent the fitting results.

**Figure 5** A numerical simulation of nanowire excitation using a laser beam. (a) Schematic of the setup: silver nanowires (diameter 350 nm) coated with silica (thickness 8 nm) are placed on a silica substrate and illuminated using a laser beam perpendicular to the substrate surface with an electric field either perpendicular or parallel to the nanowires. Images (b), (c) and (d) are the electric field intensities for single, double and triple nanowires illuminated using parallel polarization. Images (e), (f) and (g) are the electric field intensities for single, double and triple nanowires illuminated using perpendicular polarization.

**Figure 6** Plot of the average electric intensity on the nanowire surface (a) and at the hot spot (b) for single, double and triple nanowires versus the laser wavelength. The incident laser beam was polarized perpendicular to the nanowire.

**Figure 7** (a) Average electric field intensity for double nanowires versus the gap between them. (b) An illustration of a dipole antenna composed of coupled nanowires. (c) The far-field emission pattern of a dipole.

**Figure:**

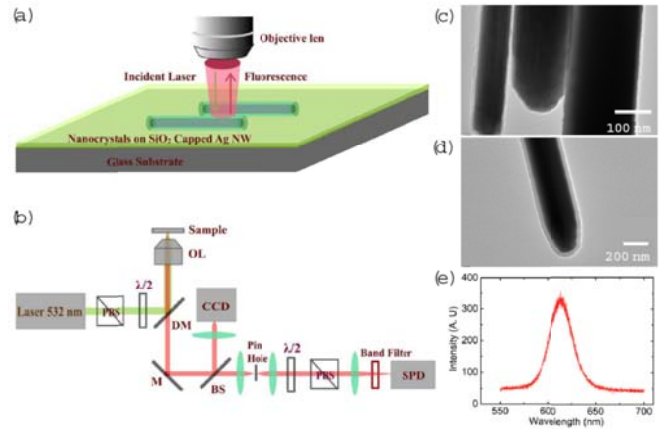

**Figure 1** (a) The pump laser illuminates the sample from the top, and the quantum dot emissions are collected using the same objective lens. CdSe/ZnS quantum dots and silver nanowires are deposited onto a glass substrate. (b) Schematic of the experimental setup. Both (c) and (d) are the TEM images of the silica capped nanowires with silica thicknesses of 8 nm and 35 nm, respectively. (e) The photoluminescence of the CdSe/ZnS quantum dots.

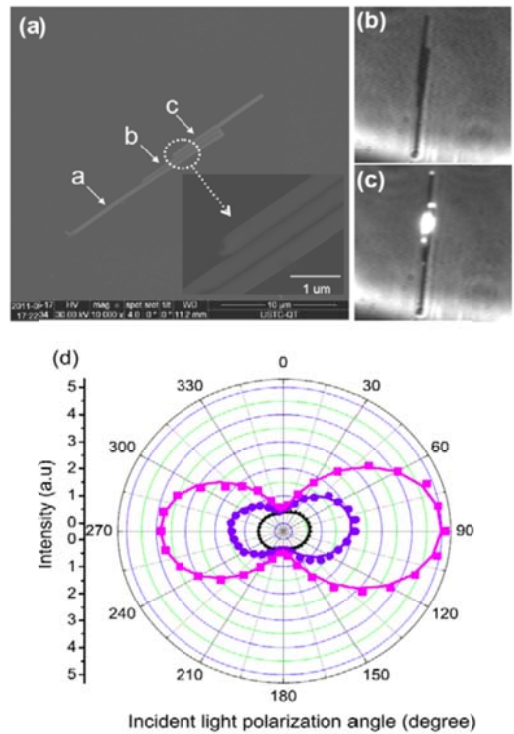

**Figure 2** (a) SEM and (b, c) CCD images of the triply aligned silver nanowires capped with 8 nm silica. (d) The fluorescence intensity as a function of the excitation laser polarization. The fluorescence curves at sites a (dark diamond), b (purple circle), and c (red square) as shown in Figure (a). The lines in (d) are the fitting curves using equation (1).

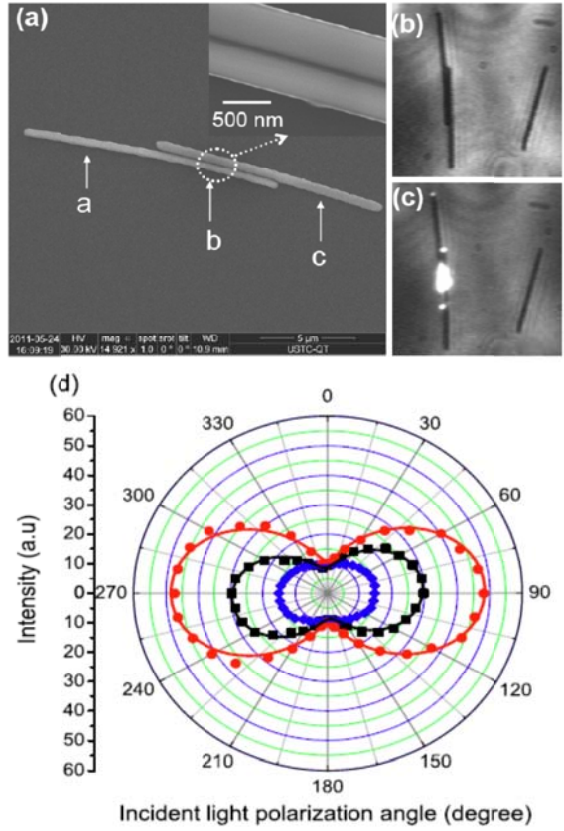

**Figure 3** (a) The SEM and (b, c) CCD images of doubly aligned silver nanowires capped with 35 nm silica. (d) The fluorescence intensity as a function of excitation polarization. The fluorescence curves at sites a (full square), b (full circle), and c (full diamond) from the three nanowires in Figure (a). The lines in (d) are the best fitting sinusoidal function.

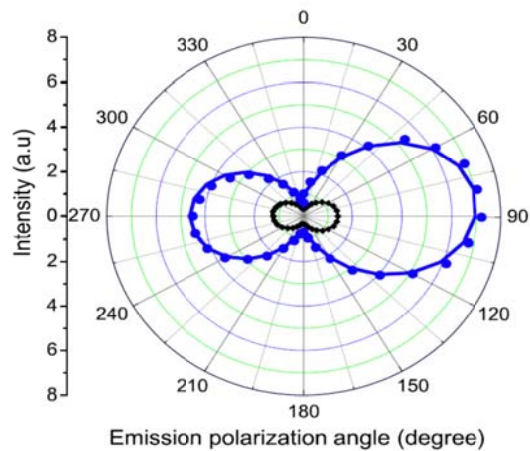

**Figure 4** The emission polarization of the fluorescence detected at sites a and c (Figure 2a) for nanowires capped with 8 nm silica. The full diamonds (curve a) and circles (curve c) correspond to sites a and c, respectively. These points represent the experimental data, while the lines represent the fitting results.

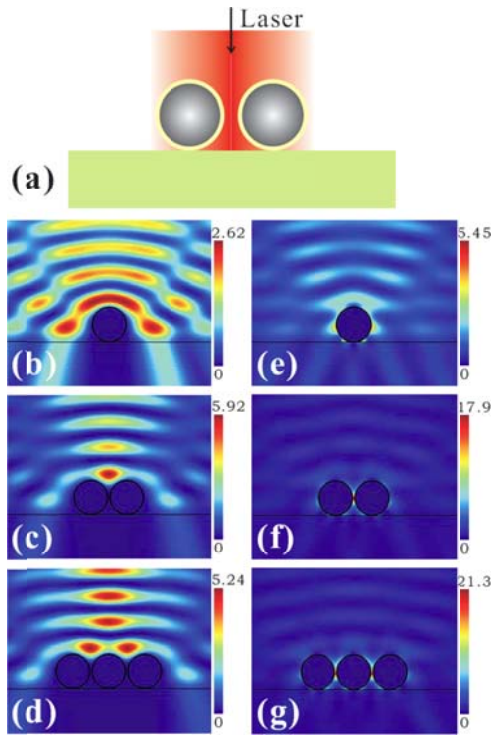

**Figure 5** A numerical simulation of nanowire excitation using a laser beam. (a) Schematic of the setup: silver nanowires (diameter 350 nm) coated with silica (thickness 8 nm) are placed on a silica substrate and illuminated using a laser beam perpendicular to the substrate surface with an electric field either perpendicular or parallel to the nanowires. Images (b), (c) and (d) are the electric field intensities for single, double and triple nanowires illuminated using parallel polarization. Images (e), (f) and (g) are the electric field intensities for single, double and triple nanowires illuminated using perpendicular polarization.

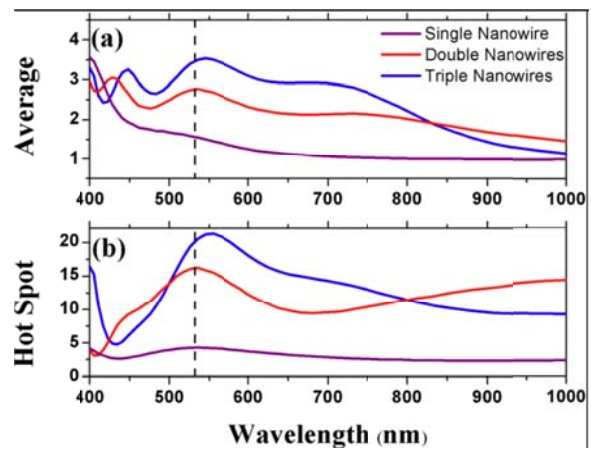

**Figure 6** Plot of the average electric intensity on the nanowire surface (a) and at the hot spot (b) for single, double and triple nanowires versus the laser wavelength. The incident laser beam was polarized perpendicular to the nanowire.